\documentstyle[twocolumn,aps,epsfig]{revtex}
\tighten
\begin{document}
\twocolumn[\hsize\textwidth\columnwidth\hsize\csname @twocolumnfalse\endcsname
\title{Wetting of Argon on CO$_2$}
\author{Giampaolo Mistura, Francesco Ancilotto, Lorenzo Bruschi and Flavio Toigo} 
\address{Istituto Nazionale per la Fisica della Materia and 
Dip.to di Fisica G.
Galilei \\ Universit\'{a} di Padova, via Marzolo 8, 35131 Padova, Italy}
\maketitle
\begin{abstract}
We have studied the wetting transition of Ar 
adsorbed on solid CO$_2$ by means of 
high-precision adsorption isotherms measured with a 
quartz microbalance. We observe triple-point wetting.
At variance with many theoretical studies based on a model adsorption potential, 
which predict for this system a genuine prewetting transition around 
$100K$, we find that a detailed density-functional calculation 
employing a more realistic adsorption 
potential leads to triple-point wetting of Ar on 
CO$_2$, in good agreement with the experiment.
\bgroup\draft
\egroup
\end{abstract}
]

Recently, there has been a tremendous progress in the experimental
investigations of wetting phenomena. For the first time, conclusive evidence
for the first-order nature of the wetting transition, as originally
predicted by Cahn\cite{cahn} long time ago, together with its accompanying
prewetting jumps away from the coexistence line, has been obtained in
several different systems: quantum liquid films physisorbed on heavy alkali
metals\cite{taborek}, complex organic liquids\cite{bonn}, near critical
liquid mercury\cite{fisher} and binary liquid crystal mixtures\cite{lucht}.

The first evidence of such a transition came however from mean-field
numerical calculations of Ar adsorbed on solid CO$_2$ carried out by Ebner
and Saam\cite{saam}. Several groups have subsequently studied this same
model system with different numerical and analytical techniques and have
essentially confirmed the original picture\cite{all,monson}. The most recent
and accurate calculation predicts a prewetting transition of Ar on CO$_2$
around 105$^\circ K$ \cite{monson}. 
It is curious that despite such a keen theoretical
interest, the wetting properties of Ar on solid CO$_2$ have never been
investigated in a real experiment, to the best of our knowledge.

\begin{figure}[tbh]
\centerline{\psfig{file=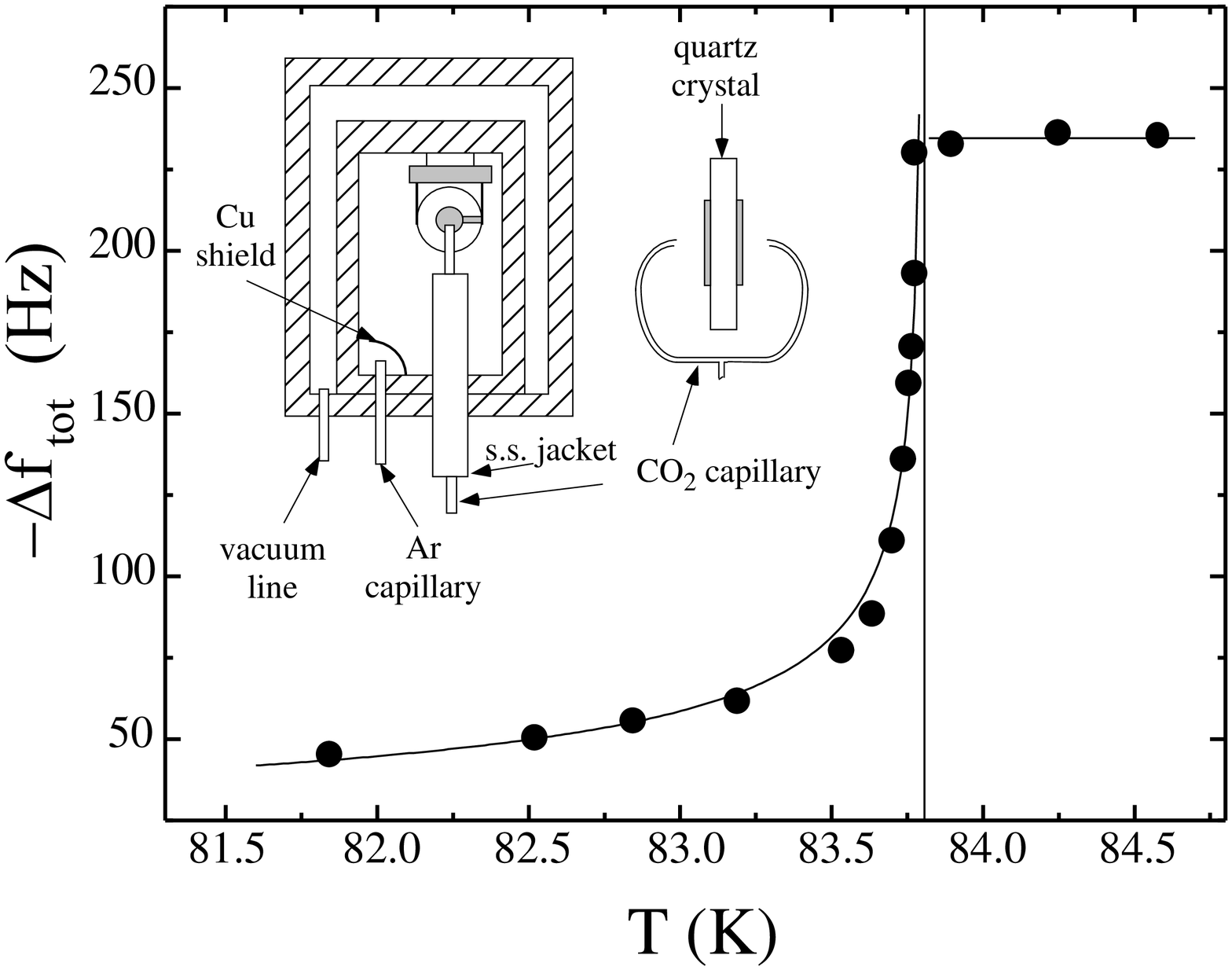, width=7cm, angle=0}}
\caption{ Total film thickness of Ar on CO$_2$ as a function of T. The
vertical line indicates T$_t$. Inset: schematic drawing of the sample cell
housing the quartz crystal and enlargement of the CO$_2$ evaporator.}
\label{fig1}
\end{figure}

We have thus decided to study such a system with a quartz microbalance
technique. This is a powerful quantitative probe in which the resonance
frequency of an AT-cut crystal excited in its shear mode depends, among
other things, on the mass of the film adsorbed on the metal electrodes of
the quartz plate. To drive the crystal to its resonance frequency we have
used an FM-technique described in detail elsewhere\cite
{bruschi}. The quartz plate employed in this study has a fundamental
resonance frequency of $5MHz$ with optically polished gold electrodes. The
frequency shift caused by the adsorption of a monolayer of liquid Ar,
$-\Delta f_m$, 
corresponds to approximately $6Hz$. The resolution is of about $\pm 0.05Hz$
with an excitation power dissipated onto the crystal of less than $50nW$.
The AT-plate is housed in a double-wall OFHC copper cell (see Fig. (\ref
{fig1})) to reduce thermal gradients. The entire set-up is attached to the
cold flange of an homemade liquid nitrogen cryostat. The maximum temperature
difference between the top and bottom parts inside the inner copper cell 
is estimated to be less than $1\,\mu K$. The
temperature stability is better than $0.5\,mK$. The Ar gas is admitted
into the cell through a thin capillary very well heat sunk to the base of
the copper cells. A copper foil, glued to the bottom of the inner cell,
faces the extremity of the Ar capillary to shield the quartz crystal from
the incoming warm vapor which is slowly admitted into the cell during the
measurements. 

The CO$_2$ substrate is evaporated at low temperature ($\approx 85^{\circ }K$
) directly onto the two opposite gold electrodes of the quartz crystal
through a different capillary also connected to the gas system. At the
temperature of evaporation, the vapor pressure of CO$_2$ is 
$\approx 1\mu Torr.$ CO$_2$ is thus admitted to the cell through a copper
capillary mounted inside a stainless steel tube, 
soldered to the base of the inner cell (see Fig. (\ref{fig1})). The copper
capillary, wrapped by a resistive wire, is
soldered to the opposite end of the s.s. tube . In this way, it is
possible to heat up the CO$_2$ capillary to a temperature much higher than
that of the quartz and thus avoid its blocking during the evaporation. The
capillary ends in two curved pieces placed in front of the two faces of the
quartz crystal (see Fig. (\ref{fig1})). 
By slowly admitting high-purity CO$_2$ gas through this heated
capillary it is then possible to evaporate CO$_2$ films with a mass
corresponding to thicknesses ranging from 100 to 1000 layers over a period
of about 2-10 minutes. 

The wetting transition of Ar on CO$_2$ has been experimentally investigated
by means of adsorption isotherms. These are determined by slowly admitting
small amounts of high-purity Ar gas into the sample cell, kept at a
constant temperature. The resonance frequency and the pressure inside the
sample cell are continuously monitored with a personal computer. Saturation
is reached when $P$ remains unchanged after further admission of a small
amount of gas. 

Fig.(\ref{fig1}) shows the results of a set of adsorption isotherms taken
across the bulk triple-point of Ar on gold preplated with more than 100
equivalent layers of CO$_2.$ The vertical axis reports the total frequency
shift $-\Delta f_{tot}$  measured at pressures between vacuum and coexistence 
vapor pressure and corrected by vapor contributions in the standard way
\cite{taborek,bruschi}.
Since the film thickness is proportional to the corrected
frequency shift, the maximum thickness of the Ar liquid films is found to
remain practically constant over the investigated temperature range, as the
horizontal line clearly shows. Finally, on bare gold $-\Delta f_{tot}$ is
only $5\%$ smaller than on this CO$_2$ plating. This difference is likely to
be caused by the rough CO$_2$ substrate evaporated at low temperature which
makes the active surface area of the microbalance larger than that of the
gold electrodes. However, such a small effect indicates that the roughness
of this CO$_2$ substrate does not play a significant role in our
measurements. 

\begin{figure}[tbh]
\centerline{\psfig{file=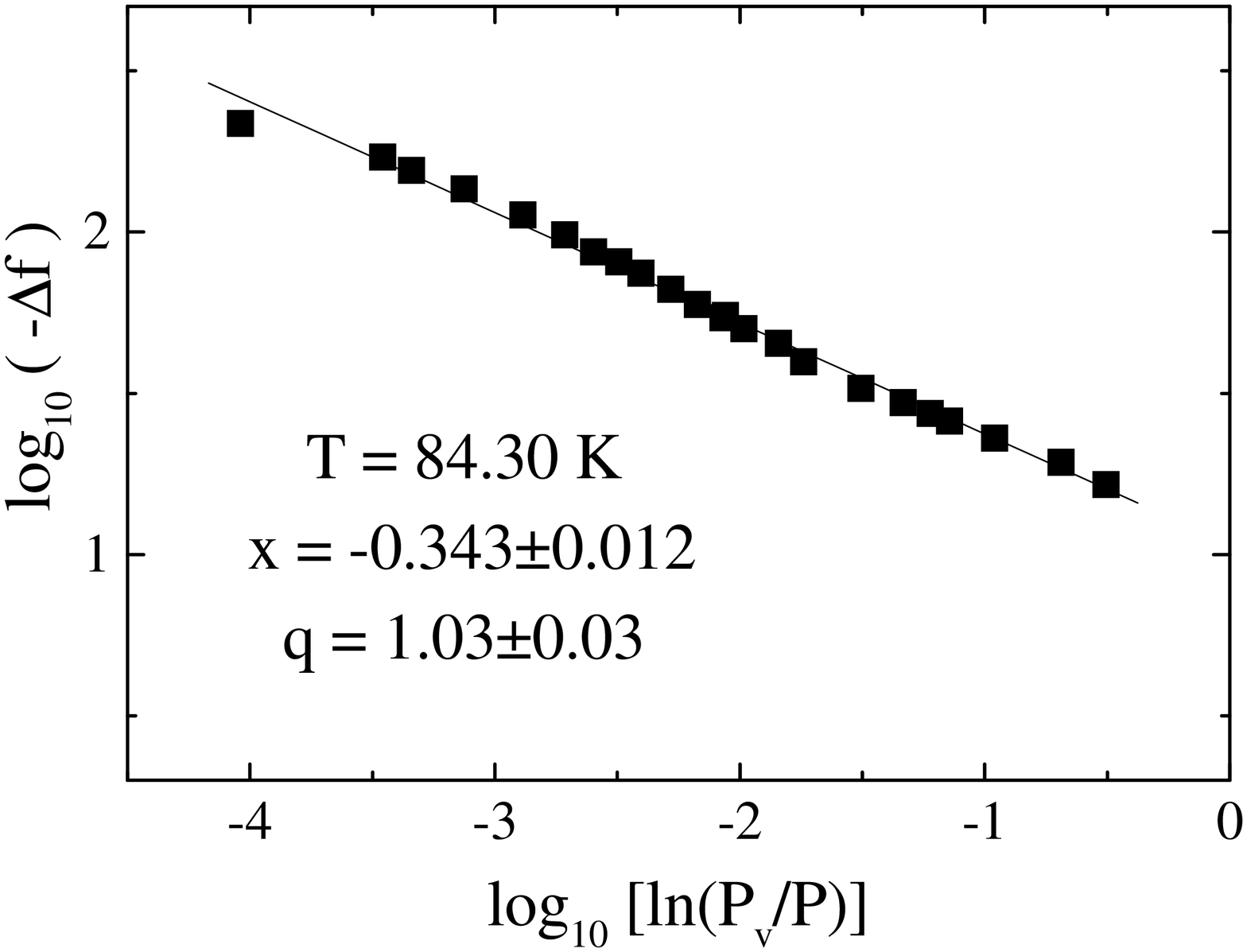, width=5.5cm, angle=-90}}
\caption{ Determination of the growth exponent $x$ of Ar on CO$_2$ }
\label{fig2}
\end{figure}

In the solid phase, $-\Delta f_{tot}$ is found to grow continuously as $T$
approaches $T_t$. We have also taken isotherms on gold preplated with
different CO$_2$ coverages ranging from 200 to about 1000 equivalent layers.
These measurements agree well with the results of Fig. (\ref{fig1})
apart from an increased frequency noise and a larger value of $-\Delta
f_{tot}$ observed at the higher coverages, probably caused by rougher
surfaces. Similar behavior, seen in all physisorbed systems characterized 
by strong
substrates, is typical of triple-point wetting. We have thus fitted the
points below $T_t$ in Fig. (\ref{fig1}) according to the power law expected
for this phenomenon\cite{krim}, 
$-\Delta f_{tot}=[AT(T_t-T) ^x$, where 
$A=\frac{-\Delta f_m^{-3}}{\alpha _{Ar-CO_2}
}\left[ \left( \frac 1{P_v}\frac{dP_v}{dT}\right) _{sol}-\left( \frac 1{P_v}
\frac{dP_v}{dT}\right) _{liq}\right] ,$ $\alpha _{Ar-CO_2}$ being the
adsorbate-substrate Hamacker constant, $-\Delta f_m=6Hz$ the frequency shift
corresponding to the adsorption of an Ar monolayer and $x=-1/3$ for van der
Waals interactions. The results of the nonlinear least square fit are: $
T_t=83.80\pm 0.02K,$ in very good agreement with the tabulated bulk
triple-point of Ar, $T_t=83.806$ $K$ , $x=-0.337\pm 0.008$ and $A=1.5\times
10^{-5}$ $K^{-2}Hz^{-3}$. We have also analyzed the film growth of the
isotherms taken above $T_t.$ Under the assumption of complete wetting at
coexistence, the film thickness is predicted to grow according to the
Frenkel-Halsey-Hill relation\cite{krim}, $-\Delta f=-\Delta f_m\left[ \frac T
{\alpha _{Ar-CO_2}}\ln \left( \frac{P_v}P\right) \right] ^{-1/3}$ where the
exponent $-1/3$ reflects again the van der Waals nature of the interaction.
In Fig.(\ref{fig2}) the quantity 
$\log _{10}\left( -\Delta f\right) $ is plotted as a
function of $\log _{10}\left[ \ln (P_v/P)\right]$. As expected, near
saturation the experimental points lie on a reasonable straight line. The
slope of the linear fit is $-0.343\pm 0.012,$ in good agreement with the FHH
equation. From the intercept $q$ we have also estimated a value for $\alpha
_{Ar-CO_2}=19000K\AA ^3\pm 20\%,$ in reasonable agreement with the van der
Waals tail used by Ebner and Saam\cite{saam}. A similar value, although
affected by a larger error, has been deduced from the parameter $A$ of the
the previous fit.

Motivated by these experimental findings, which are in disagreement 
with the
previous theoretical predictions of a prewetting transition in this system,
we have reexamined the basis for such predictions and found that they are
flawed in one of their basic premises, i.e. in the form of the
fluid-substrate interaction potential. Common to these theoretical studies
is in fact the assumption, as made by Ebner and Saam \cite{saam}, that the
Ar-CO$_2$ interaction is represented by a (12-6) Lennard-Jones potential
whose hard-core diameter and well depth are deduced, through simple
combining rules\cite{steele}, from those of Ar-Ar and CO$_2$-CO$_2$. 
An integration of the (12-6) potential over a continuum of 
CO$_2$ substrate atoms produces a (9-3) adsorption potential, hereafter called
Ebner-Saam (ES) potential, which has been used in almost all the theoretical
calculations published so far\cite{all,monson}.

By comparing this potential with a more realistic one, which we compute by
explicitly taking into account the structure of the CO$_2$ substrate, as shown below,
we find that the ES potential underestimates the real interaction between Ar
atoms and solid CO$_2$. Our density functional calculations
encompassing the newly calculated adsorption potential support the
experimental evidence reported above of triple-point wetting.

In our density functional calculation the free energy of the
fluid is written in term of the density $\rho (\vec{r})$ of the fluid as: 
\begin{eqnarray}
F[\rho ] &=&F_{HS}[\rho ]+\int \rho (\vec{r})V_s(\vec{r})d\vec{r} \nonumber \\
&& +{\frac 12}\int \int \rho (\vec{r})\rho (\vec{r}^{\, \prime })u_a(|\vec{r}-
\vec{r}^{\, \prime }|)d\vec{r}d\vec{r}^{\, \prime }  \label{energy}
\end{eqnarray}

Here $F_{HS}$ is a non-local free-energy functional for the inhomogeneous
hard-sphere reference system, 
$V_s(\vec{r})$ is the external adsorption
potential due to the surface, 
while the third term is the usual mean-field
approximation for the attractive part of the fluid-fluid intermolecular
potential, $u_a$. $F_{HS}$ is evaluated according to the theory of 
Ref.\cite{rosinberg}, which has been found to give an
accurate description of inhomogeneous fluids even in situations
where strong density variations occur, as in the presence of hard walls or
strongly attractive substrates. 
Following Ref.\cite{monson,tarazona}
we describe the attractive part of the interparticle potential 
as an effective interaction 
\begin{eqnarray}
u_a(r) &=&0\,\,\,\,\,,\,\,\,r\le \lambda ^{1/6}\sigma   \nonumber \\
&=&4\epsilon \{\lambda (\sigma /r)^{12}-(\sigma /r)^6\}\,\,\,\,\,\,,\,\,\,
r>\lambda ^{1/6}\sigma
\end{eqnarray}

We determine the free parameters $\lambda $ and $\sigma $ 
(the HS diameter) for each
temperature by requiring that the experimental values of liquid and vapor
densities at coexistence, $\rho _l$ and $\rho _v$, are reproduced for the bulk
fluid. In order to correctly describe the bulk phase diagram, it is
important to verify that the thermodinamic equilibrium conditions
are satisfied. This is done by means of a Maxwell (equal-area)
construction in the $P-\rho $ plane.

CO$_2$ has a unique crystalline phase at low temperature and pressure \cite
{keesom}. Thin solid films of CO$_2$ are known to have crystalline
structure\cite{weida} when grown at $T>80\,^{\circ }K$ and in a wide
range of growth rates, including those realized in our experiment. Moreover,
it is found experimentally \cite{weida} that the exposed surface is almost
invariably the (001) face of the crystal.

\begin{figure}[htb]
\centerline{\psfig{file=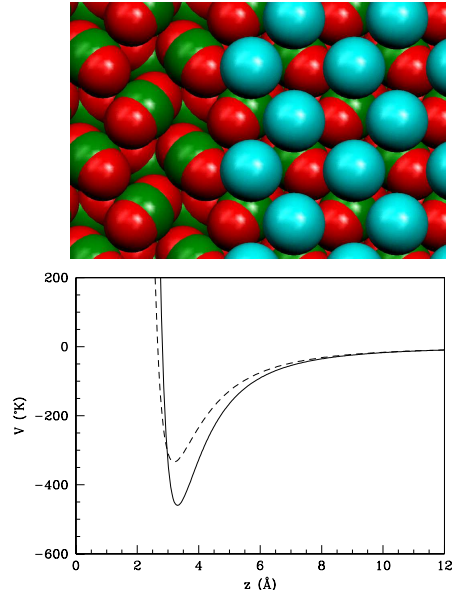, width=7cm, angle=0}}
\caption{ Upper part: top view of the CO$_2$ surface. The left part shows
the clean, ideal surface: C atoms and O atoms are represented by the green
and red balls respectively. The right part shows the surface covered with a
monolayer of Ar atoms (blue balls), in the positions as explained in the
text. Lower part: fluid-substrate laterally-averaged potential. The dashed
line shows the ES potential, while the solid line is our calculated
potential. See text for the definition of the $z=0$ planes.}
\label{fig3}
\end{figure}

Accurate {\it ab initio} potentials are available\cite{marshall}, which
describe the interaction between a CO$_2$ molecule and an Ar atom.
Due to the nature of the bonding in CO$_2$ solid (van der Waals +
electrostatic forces) no appreciable electron charge redistribution is
expected between the molecules in the solid. We may thus calculate the total
Ar-surface potential by directly summing the two-body interactions between
an Ar atom and the CO$_2$ molecules in the crystal, 
assuming that the exposed surface is the ideal (001).
The resulting potential energy surface as a function of
the lateral position of the Ar atom exhibits large variations, with
differences in binding energies for different adsorption sites of more than 
$500\,^{\circ }K$. In particular, in the $c(2\times 2)$ surface unit cell of
the clean (001) surface there are two (non-equivalent) 
adsorption sites with
very large ($\sim 1000^\circ K$) binding energies. We then make the 
assumption that at the low temperatures investigated here, at least one Ar
monolayer is adsorbed on these sites 
as a solid-like layer. We determine the Ar equilibrium
positions by energy optimization, neglecting however any relaxation of the
CO$_2$ surface. The structure of the adsorbed monolayer is shown in
the upper part of Fig. (\ref{fig3}). We
finally average laterally the interaction potential between Ar and
this Ar-preplated surface, to get the adsorption potential $V_s(z)$ to be
used in our calculations ($z$ is the 
coordinate normal to the surface plane). 
This is shown in the lower part of Fig. (\ref{fig3}),
where it is compared with the original ES potential.    
The origin of the z axis  
is taken at the position of the outmost 
layer of C atoms in the case of the ES potential, while it is 
taken at the position of the solid Ar monolayer
for our calculated $V_s(z)$.
Note that the net adatom-surface potential is deeper than the ES
potential even if the CO$_2$ surface is screened by the solid Ar layer.
This is a consequence of the large corrugation of the CO$_2$ surface:
the structure and orientation of the CO$_2$ molecules on the surface plane
(see Fig.(\ref{fig3})) allow the Ar monolayer to be easily
accomodated but still leaving space 
for additional Ar atoms to come close to the surface, thus
experiencing the attractive part of the interaction.

The equilibrium Ar density profile $\rho (z)$ is determined by direct
minimization of the functional (\ref{energy}) with respect to density
variations. In practice we fix the coverage $\Gamma =\int dz \rho (z)$ and
solve iteratively the Euler equation $ \mu = \delta F/\delta \rho(z)$, where
the value of the chemical potential $\mu $ is fixed by $\Gamma $.

\begin{figure}[tbh]
\centerline{\psfig{file=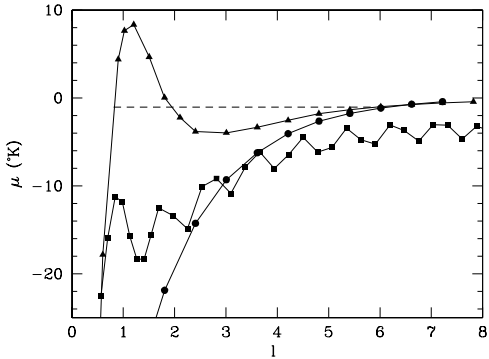, width=6.5cm, angle=0}} 
\caption{ Calculated adsorption isotherms of Ar in the fluid phase. 
Triangles: $T=105^{\circ }K$, as
calculated with the ES potential. Dots: $T=105^{\circ }K$, with the new
potential. Squares: $T=85^{\circ }K$, with the new potential. 
}
\label{fig4}
\end{figure}

We have considered two isotherms, one at $T=105\,^{\circ }K$, which has been
studied in most detail in Refs.\cite{monson}, and one at 
$T=85\,^{\circ }K$, i.e. just above the triple point $T_t$. We
report our results in Fig. (\ref{fig4}), where the chemical potential,
measured with respect to its value at coexistence, is
plotted vs. the film thickness. The latter is expressed in nominal layers 
$l=\rho_l^{-2/3}\int_0^\infty [\rho (z)-\rho _v]dz$.
The dots show the calculated
points at $T=105^{\circ }K$, where one observes a smooth increase of
coverage with chemical potential as the saturated liquid is approached from
below. At variance with the results of Ref.\cite{monson}, we do
not observe any sign of a prewetting transition. We show for comparison in the
same figure the results obtained, at the same temperature, by using the ES
potential (triangles). 
The large oscillation with positive values of $\mu $ is
representative of a thin film-thick film prewetting transition, in agreement
with previous calculations using the same potential\cite{saam,all,monson}. A jump
in the coverage from $l\sim 0.8$ to $l\sim 6$ can be readily obtained from our
data by means of a Maxwell equal-area construction (dashed line). 

The calculated points at $T=85^{\circ }K$ are shown with squares. 
A sequence of oscillations occurs in this case, whose period is
one nominal layer, corresponding to 
the condensation of successive Ar monolayers ("layering" transitions).

Additional theoretical support to triple-point wetting of Ar on CO$_2$
can be obtained from a simple but surprisingly accurate heuristic model \cite
{cheng} where the energy cost of forming a thick film is compared
with the benefit due to the gas-surface attractive interaction, 
\begin{eqnarray}
(\rho _l-\rho _v)\int_{z_0}^\infty V_s(z)dz=-2\gamma
\label{criterion}
\end{eqnarray}

Here $\rho _l$ and $\rho _v$ are the densities of the adsorbate liquid and
vapor at coexistence, $\gamma $ is the surface tension of the liquid and $z_0$
is the equilibrium distance of the potential $V_s$. By using for $V_s$ our
calculated potential and the
experimental values for $\rho _l(T)$, $\rho _v(T)$ and $\gamma (T)$,
we find that Eq.(\ref{criterion})
is satisfied at $T_W=1.03\,\,T_t$. 

In conclusions, our
experimental results suggest triple-point
wetting of Ar on solid CO$_2$, in contrast to the many previous theoretical
investigations. To clarify this finding, we have
implemented density functional calculations. A crucial ingredient
is a realistic form for the fluid-surface potential, which 
is obtained by explicitly considering the microscopic 
structure of the surface of solid CO$_2$ and the anisotropy of the Ar-CO$_2$ 
interaction.
The result is a stronger potential than originally estimated by Ebner and Saam
by summing isotropic Ar-CO$_2$ LJ interactions over a continuum substrate.
The increased binding of the surface changes dramatically the Ar
growth mode, which now results in a continuous (layer-by-layer) film growth 
rather than exhibiting a prewetting transition. 
The results of our calculations are 
consistent with a wetting temperature close to the triple point, in
agreement with our quartz microbalance measurements.

We thank Stefano Sitran and Gilberto Schiavon for the
preparation of the quartz crystals used in this work.

\end{document}